\documentclass[sigconf]{acmart}

\AtBeginDocument{%
  }

\setcopyright{acmlicensed}
\copyrightyear{2025}
\acmYear{2025}
\acmDOI{XXXXXXX.XXXXXXX}

\acmConference[Conference acronym 'XX]{Make sure to enter the correct
  conference title from your rights confirmation email}{June 03--05,
  2025}{Woodstock, NY}
\acmISBN{978-1-4503-XXXX-X/18/06}

\usepackage{multirow}
\usepackage{graphicx}
\usepackage{bm}
\usepackage{svg}
\begin{document}

\title{Explicit Uncertainty Modeling for Video Watch Time Prediction}


\author{Shanshan Wu}
\affiliation{%
  \institution{Kuaishou Technology}
  \city{Beijing}
  \country{China}}
\email{wushanshan03@kuaishou.com}

\author{Shuchang Liu}
\affiliation{%
  \institution{Kuaishou Technology}
  \city{Beijing}
  \country{China}}
\email{liushuchang@kuaishou.com}

\author{Shuai Zhang}
\affiliation{%
  \institution{Kuaishou Technology}
  \city{Beijing}
  \country{China}}
\email{zhangshuai09@kuaishou.com}

\author{Xiaoyu Yang}
\affiliation{%
  \institution{Kuaishou Technology}
  \city{Beijing}
  \country{China}}
\email{yangxiaoyu@kuaishou.com}

\author{Xiang Li}
\affiliation{%
  \institution{Kuaishou Technology}
  \city{Beijing}
  \country{China}}
\email{xiangli44@kuaishou.com}

\author{Lantao Hu}
\affiliation{%
  \institution{Kuaishou Technology}
  \city{Beijing}
  \country{China}}
\email{hulantao@kuaishou.com}

\author{Han Li}
\affiliation{%
  \institution{Kuaishou Technology}
  \city{Beijing}
  \country{China}}
\email{lihan08@kuaishou.com}

\renewcommand{\shortauthors}{Shanshan et al.}

\begin{abstract}
In video recommendation, a critical component that determines the system's recommendation accuracy is the watch-time prediction module, since how long a user watches a video directly reflects personalized preferences.
One of the key challenges of this problem is the user's stochastic watch-time behavior.
To improve the prediction accuracy for such an uncertain behavior, existing approaches show that one can either reduce the noise through duration bias modeling or formulate a distribution modeling task to capture the uncertainty.
However, the uncontrolled uncertainty is not always equally distributed across users and videos, inducing a balancing paradox between the model accuracy and the ability to capture out-of-distribution samples.
In practice, we find that the uncertainty of the watch-time prediction model also provides key information about user behavior, which, in turn, could benefit the prediction task itself.
Following this notion, we derive an explicit uncertainty modeling strategy for the prediction model and propose an adversarial optimization framework that can better exploit the user watch-time behavior.
This framework has been deployed online on an industrial video sharing platform that serves hundreds of millions of daily active users, which obtains a significant increase in users' video watch time by 0.31\% through the online A/B test. 
Furthermore, extended offline experiments on two public datasets verify the effectiveness of the proposed framework across various watch-time prediction backbones.
\end{abstract}



\keywords{Recommender System, Stochastic Modeling, Quantile Regression}


\maketitle

\begin{figure*}[t]
    \centering
    \includegraphics [width=\textwidth]{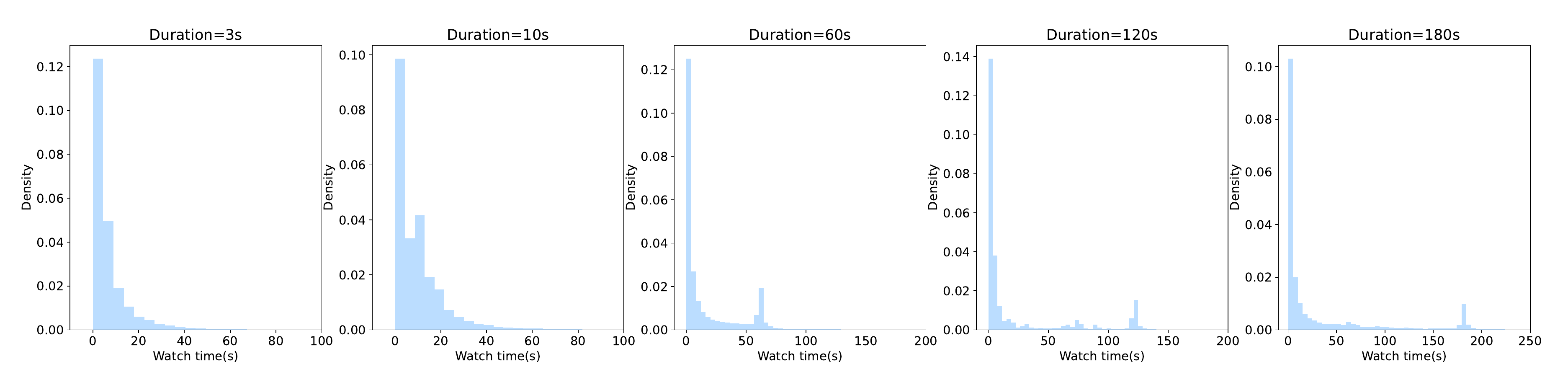}
    \caption{The distribution of watch time in different subsets grouped by duration. Data is collected from an online industrial video sharing platform.}
    \label{fig: watchtime_distribution}
\end{figure*}

\section{Introduction}\label{sec: intro}

Recommender systems~\cite{koren2009matrix,aggarwal2016recommender} serve as the main building block in a wide range of web services and aim to select the best items according to users' interests.
In conventional recommendation settings in e-Commerce, news, or advertising, users provide binary feedback that represents positive or negative interactions, and the problem is often formulated as a click-through rate prediction~\cite{zhou2018deep} or a learning-to-rank problem~\cite{liu2009learning}.
In contrast, video recommendation formulates a special recommendation task in which users provide feedback as watch time~\cite{covington2016deep}, a continuous value that describes a fine-grained magnitude of user preference.
In this scenario, the accuracy of the watchtime prediction module directly determines the performance of the recommender system and the user experience.

Existing evidence has shown that the watch-time prediction problem is more complicated than a simple regression task. 
On one hand, the semi-closed prediction interval (i.e. the watch time could range from zero to infinity if allowing repeated watching) potentially generates a skewed label distribution. 
On the other hand, the watch-time distribution of users could be arbitrarily complex. 
Figure \ref{fig: watchtime_distribution} shows that the watch-time distribution could vary across different video lengths (called duration) and is highly skewed towards the short-play region compared to complete play and repeated play. 
This phenomenon can be mainly attributed to two factors:
First, there are deterministic factors, such as duration bias~\cite{zhan2022deconfounding, zhao2024counteracting}, exposure bias~\cite{abdollahpouri2020multi}, popularity bias~\cite{wei2021model, zhang2021causal}, etc. 
Second, there are noises resulting from non-deterministic factors, such as accidental device malfunction, intentional re-play, leaving the device unattended or data stream delays.
As a solution, several distribution modeling methods~\cite{lin2023tree,sun2024cread,lin2024conditional} are proposed to learn potentially arbitrary distribution patterns with the existence of deterministic biases rather than the exact watch-time value prediction.

In general, most studies agree that the watch-time label distribution does not necessarily follow a simple Gaussian or a multipeak mixture. This indicates that a flexible distribution modeling method is required to handle various user watch-time patterns.
Representative solutions include but are not limited to quantile prediction~\cite{zhan2022deconfounding} and ordinal regression~\cite{sun2024cread}.
The quantile prediction method empirically records the quantiles of the watch-time distribution, and the framework learns to predict the quantile level instead of the watch-time to deal with arbitrary user patterns.
Ordinal regression learns a model with multiple prediction heads for each quantile and predicts the probabilities of reaching each quantile, which provides more stable performance in practice.
These distribution modeling methods aim to improve the model's ability to capture the stochastic user watch time even if the distribution is complicated, hard to predict, and far from the expectation.

\begin{figure}
    \centering
    \includegraphics[width=\linewidth]{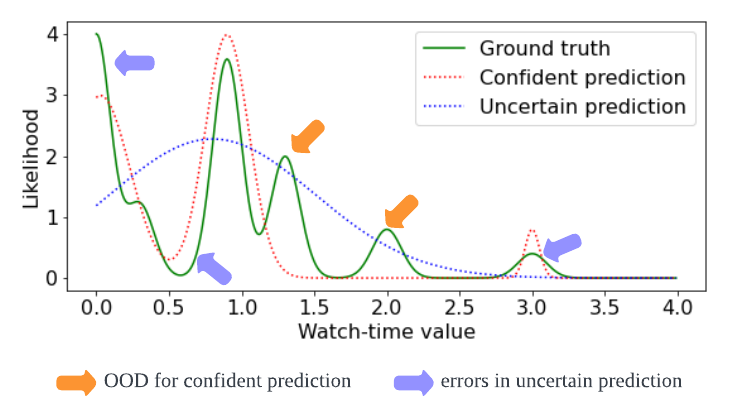}
    \caption{The uncontrolled uncertainty paradox. The confident model may capture part of the ground-truth behavior but is more likely to see out-of-distribution errors and the uncertain model may capture out-of-distribution samples but reduces prediction accuracy for existing samples.}
    \label{fig: ood_sample}
\end{figure}

Ideally, the distribution modeling models should become more and more accurate with the increasing data support.
However, in practice, users may exhibit dynamic, diverse, subtle, or even confusing behaviors from time to time, which continuously generates out-of-distribution samples and challenges the model's accuracy.
As a consequence, the optimization process generates an \textit{uncontrolled uncertainty paradox}. 
We present an intuitive example of this paradox in Figure \ref{fig: ood_sample}. 
On one end, the model has to keep a high uncertainty on the predicted distribution in order to increase the chance of hitting the ground-truth watch-time label, especially for out-of-distribution samples (e.g. hard-to-predict cases as shown by blue dashed line in Figure \ref{fig: ood_sample}).
However, over-amplifying the magnitude of model's uncertainty would induce insufficient confidence in the prediction, reducing recommendation performance. 
On the other end, ignoring the control of the uncertainty may produce a model that overfit certain simple distribution with reduced model robustness against outliers. 
The model may effectively reduce the prediction error (e.g. for easy-to-predict cases as shown by red dashed line in Fig \ref{fig: ood_sample}), but the predicted distribution may become over-confident and encounter more out-of-distribution samples during inference.
Intuitively, we need an automatic solution that provides precise control over the uncertainty of the prediction model without sacrificing the label-capturing ability.

In this work, we propose to encapsulate the original prediction model with an extra uncertainty modeling framework, where a confidence prediction module directly predicts how likely the model's prediction of the watch-time distribution covers the observed label.
Specifically, during training, the confidence prediction module behaves as a selector between the predicted watch time and the ground truth, where samples with high certainty tend to use predicted value as output and samples with low certainty tend to use the ground truth to achieve minimized error.
Yet, there is no straightforward indicator or label for the model uncertainty and we have to find a solution that learns an explicit uncertainty model.
Additionally, we theoretically show that the new confidence-controlled prediction model would consistently reduce the confidence, merely using the standard watch-time distribution learning methods (i.e., quantile prediction~\cite{zhan2022deconfounding} and ordinal regression~\cite{sun2024cread}).
The resulting framework tends to reach a shortcut that always using the ground-truth label as reflection.
To circumvent this dilemma, we include a confidence maximization objective to ensure adversarial confidence regularization and effective learning of the prediction model.
Different from the confidence regularization in other problem (e.g., classification~\cite{devries2018learning}), the distribution modeling tasks with different methods requires different adjustment strategies.
And we derive the solutions for quantile prediction and ordinal regression methods, presenting theoretically analysis and empiricaly verification on how to achieve optimal optimization by controlling the magnitude of the confidence regularization.
The resulting framework can explicitly express the confidence of the watch-time prediction model while allowing joint optimization through the adversarial learning framework.
We summarize the contributions of our work as follows:
\begin{itemize}
    \item We address the direct control of the uncertainty of the watch-time prediction in video recommendation tasks and propose an explicit modeling framework with theoretical derivation.
    \item We conduct comprehensive offline experiments as well as online A/B test to verify the effectiveness of our solution.
    \item We further show that our proposed method is a general and flexible solution that could accommodate various existing watch-time prediction models and improve their predictive accuracies.
\end{itemize}
For the rest of the paper, we will first describe related works in section \ref{sec: related_work} and illustrate our solution method in section \ref{sec: method}.
We present the experimental support for our claims in section \ref{sec: experiments} and then conclude our work in section \ref{sec: conclusion}.

\section{Related Work}\label{sec: related_work}

\subsection{Watch-time Prediction}

Video recommender systems~\cite{davidson2010youtube,zhang2021commentary} deliver personalized content to users and are one of the most critical interfaces in many online platforms like YouTube, TikTok, and Instagram.
The watch-time metric has been regarded as the core evaluation and modeling target since it provides the most elaborate feedback on the user's engagement over a video.
An early study introduces a time-weighted logistic regression~\cite{covington2016deep} to deal with the semi-closed watch-time interval during training.
Later studies have shown that the empirical watch-time distribution is closely related to the length of the video, so several duration-based watch-time prediction methods~\cite{zhan2022deconfounding,zheng2022dvr,zhao2023uncovering, zhang2023leveraging} are proposed.
While D2Q~\cite{zhan2022deconfounding} , DML~\cite{zhang2023leveraging} and D2Co~\cite{zhao2023uncovering} directly use the duration bias to correct the prediction, DVR~\cite{zheng2022dvr} formulates a debiased metric and adversarial learning to guide the model.
Except for the studies of duration bias under single Gaussian distribution, more recent studies~\cite{tang2023counterfactual} found that the user's stochastic watch-time behavior is more than a simple distribution and could be arbitrarily complex, which poses a great challenge for the prediction task.
As a solution, several distribution modeling methods~\cite{lin2023tree,sun2024cread,lin2024conditional} are proposed to learn potentially arbitrary distribution patterns.
TPM~\cite{lin2023tree} formulates a watch-time prediction tree to allow probabilistic modeling of the watch-time in hierarchical quantiles.
CREAD~\cite{sun2024cread} models discretized watch time through ordinal regression and predicts the probability of reaching each quantile.
The ongoing work CQE~\cite{lin2024conditional} further extends the idea of distribution modeling and formulates a quantile regression problem which enables the model to capture the watch-time randomness for any quantile.
As we have discussed in section \ref{sec: intro}, these distribution modeling methods can capture the stochastic user behavior but provide no explicit modeling of the uncertainty or confidence as simple variational models, which generates the uncontrolled uncertainty paradox.

\begin{figure*}[t]
    \centering\includegraphics[width=\textwidth]{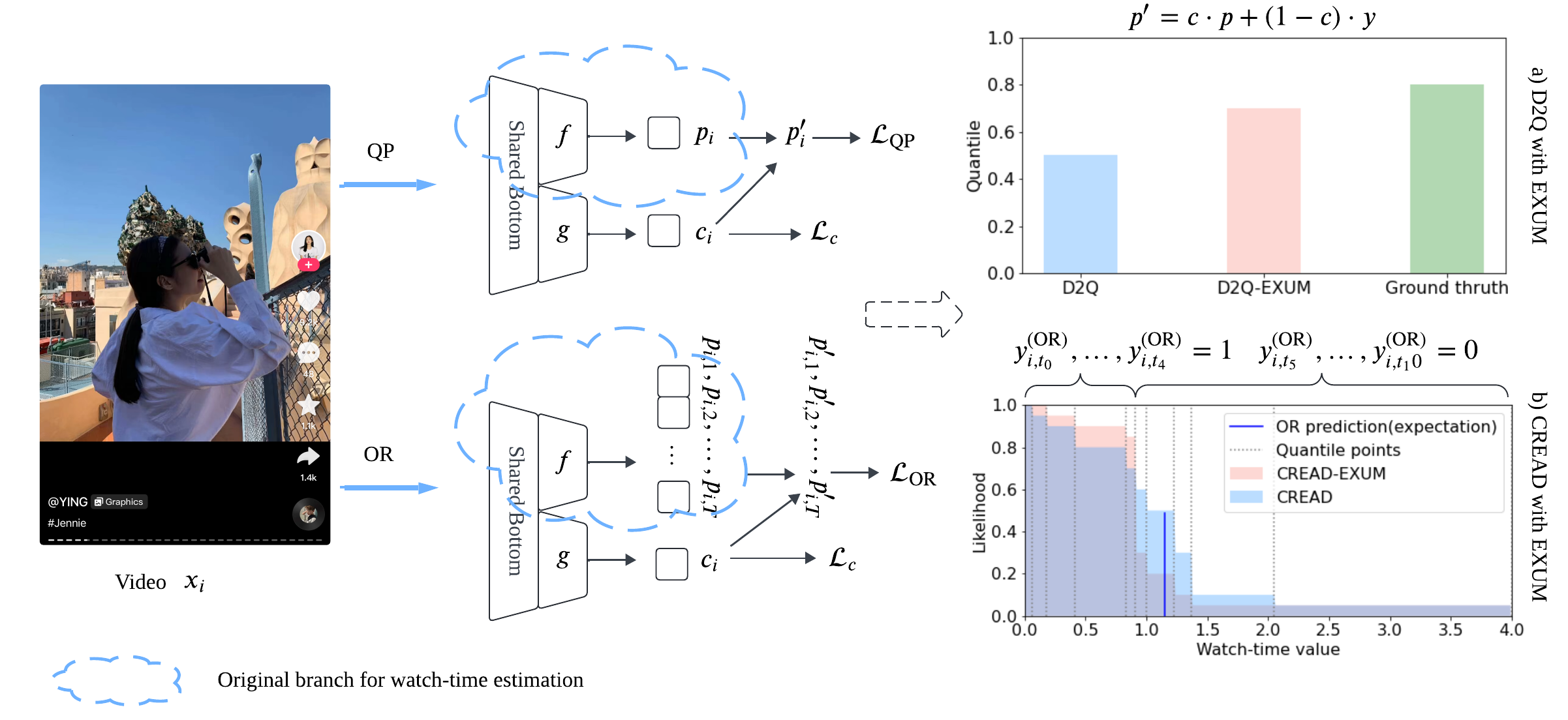}
    \caption{An instance of QP and OR problem formulation; (a) and (b): The overall learning framework of EXUM. By encapsulating the original branch for watch-time estimation with an additional branch for confidence estimation, the predictions of the model are closer to the ground truth. }
    \label{fig: overall_framework}
\end{figure*}

\subsection{Preliminary of Distribution Modeling}\label{sec: preliminary}

In this section, we present two major state-of-the-art watch-time distribution modeling methods: the quantile prediction model D2Q~\cite{zhan2022deconfounding} and the ordinal regression model CREAD~\cite{sun2024cread}.
An intuitive example of the two methods with 10 quantile points (i.e. $[10\%,\dots,100\%]$) is illustrated in Figure \ref{fig: overall_framework}.

The quantile prediction (QP) model takes an input $x$ that consists of user, video, and context information to output the quantile $p=f_\theta(x)\in[0,1]$ rather than the watch time.
In D2Q, the ground-truth label $y$ is calculated according to the watch-time value and its position in the distribution under the same video duration.
Then, it learns a standard regression task that approximates $y$ through $p$:
\begin{equation}
    \mathcal{L}_\text{QP} = \frac{1}{2}\big(p - y\big)^2
\end{equation}
This setting allows the model to model arbitrary watch-time distribution no matter the range and patterns.
One may also consider a probabilistic model that allows the nondeterministic prediction to give a more robust approximation of the distribution~\cite{zhao2023uncovering}, but we remind readers that this may require a more careful design to deal with the prediction accuracy drop in practice.

In contrast, the ordinal regression (OR) model first segments the watch-time space into empirical quantiles $t^{(0)}, t^{(1)}, \dots, t^{(N)}$ and finds the corresponding watch-time values $w^{(0)},w^{(1)},\dots,w^{(N)}$.
For example, a uniform quantization for watch-time range of $[0,10]$ would generate $N$ segments, and each segment $i$ has range $[t_{i-1},t_i]$.
In practice, we found that a better strategy is to first obtain the overall range according to the video's duration, and then adjust the watch-time distribution based on the video's popularity before obtaining the quantile points.
With the ground-truth user watch-time $\tau$, the label is set to $y^{(n)}=1$ if the corresponding watch-time has $w^{(n)}<\tau$, and $y^{(n)}=0$ otherwise, indicating whether the user has watched more than the $t^{(n)}$ portion of the video.
With this definition, the ordinal relationship states that if the label $y^{(n)}=1$ for any quantile $t^{(n)}$, then $y^{(m)}=1,\forall t^{(m)}<t^{(n)}$.
During training, the model $p$ separately learns for each quantile head with binary cross entropy:
\begin{equation}
    \mathcal{L}_\text{OR} = \sum_{n} y^{(n)} \log p^{(n)} + (1-y^{(n)}) \log (1 - p^{(n)})
\end{equation}
During inference, the expected watch-time $\mathbb{E}[\tau]$ is approximated as the weighted sum of the quantile segment length:
\begin{equation}
    \mathbb{E}[\tau] = \sum_{n>0} p^{(n)} (w^{(n)} - w^{(n-1)})
\end{equation}
As illustrated by the example in Figure \ref{fig: overall_framework}-b, the blue shade provides the probabilistic estimation $p^{(n)}$ of each quantile segment $[t^{(n-1)}, t^{(n)}]$.
The expectation is represented by the blue line with a watch-time value around 1.15 in between the quantile segment $[t^{(6)}, t^{(7)}]$.

\section{Method}\label{sec: method}

In this section, we introduce a confidence prediction module that explicitly expresses how confident it is about the watch-time prediction.
Then we illustrate the model's tendency of shortcut finding under the new modeling strategy through a theoretical analysis of the confidence degradation.
We further show that we can alleviate this problem by introducing a simple and effective adversarial learning objective, which collaboratively achieves the joint optimization of the confidence model and the watch-time prediction model.
We denote our proposed solution framework as EXplict Uncertainty Model (EXUM) and present the overall design in Figure \ref{fig: overall_framework}.

\subsection{Model-agnostic Explicit Uncertainty Model}

\textbf{Quantile Prediction:}
Assume that we have a backbone watch-time quantile prediction model $p_i = f_\theta(\bm{x}_i) \in [0,1]$ where the model is parameterized by $\theta$ (which can be any quantile prediction model similar to~\cite{zhan2022deconfounding}) and $\bm{x}_i$ denote the input that may contain information about the user, the video, and the context of the recommendation request (e.g. device, time zone, etc.)
To explicitly model the uncertainty of the model's prediction, we introduce the confidence model $c_i = g_\phi(\bm{x}_i) \in  [0,1]$ that acts as an ensemble selector between the predicted value and the ground truth during training:
\begin{equation}
    p^\prime_i = c_i p_i + (1-c_i) y_i\label{eq: quantile_prediction}
\end{equation}
where $y_i$ is the ground-truth label of the user's watch-time and $\phi$ is the parameter set of the confidence prediction model.
For implementation, one can use MLP for both $f_\theta$ and $g_\phi$, but in our practice, including a shared-bottom module between these two models is critical to guarantee the knowledge transfer from the confidence and the watch-time prediction, as illustrated in Figure \ref{fig: overall_framework}-a and Figure \ref{fig: overall_framework}-b.
Ideally, the watch-time prediction model is confident (i.e. $c_i\approx 1$) when the prediction is close to the ground truth $p_i\approx y_i$.
Otherwise, it is uncertain (i.e. $c_i\approx 0$) when the prediction is far from the ground truth.
In both cases, the combined value of $p_i^\prime$ is ensured to be close to the ground truth when the confidence model accurately represents the error.
In this sense, the relation of $p_i$ and $c_i$ is similar to that between the Gaussian distribution and its variance parameter, but $c_i$ in our case describes the `expected error' for an arbitrary distribution.

As illustrated in Figure \ref{fig: overall_framework}-a, when integrating the optimization framework we can simply substitute $p$ with $p^\prime$ and formulate the following objective functions for quantile prediction models:
\begin{equation}
    \mathcal{L}_\text{QP} = \sum_{i} \frac{1}{2}\big(p_i^\prime - y_i\big)^2 \label{eq: quantile_prediction_loss}
\end{equation}
where the label $y_i\in[0,1]$ represents the watch-time quantile under the corresponding duration (i.e. the video length) group.
Now derive the gradient of $p_i$ as:
\begin{equation}
    \frac{\partial \mathcal{L}_\text{QP}}{\partial p_i} = c_i^2(p_i-y_i)\label{eq: QR_p_gradient}
\end{equation}
which means that $p_i$ will learn faster when the model is confident (i.e., large $c_i$) or observes a large error (i.e., large $|p_i-y_i|$).
In contrast, $p_i$ will learn slower when the model is uncertain (i.e. small $c_i$ that can tolerate more errors) or observes a small error (even when the model is confident).

\textbf{Ordinal Regression:}
When adopting the ordinal regression framework like~\cite{sun2024cread}, the prediction model has $T$ quantile prediction heads, and each head $t$ outputs the probability $p_{i,t} = f_\theta(\bm{x}_i)\in[0,1]$ of the user's watch-time being more than that quantile in the distribution.
Now, as illustrated in Figure \ref{fig: overall_framework}-b, we can adopt the same strategy as Eq.\eqref{eq: quantile_prediction} only that the strategy integrates the confidence module with each of the prediction heads:
\begin{equation}
    p^\prime_{i,t} = c_i p_{i,t} + (1-c_i) y_{i,t} \label{eq: ordinal_prediction}
\end{equation}
where $c_i$ represents the confidence of the overall distribution prediction, so it applies to all the prediction heads.
Ideally, the model is confident (i.e. $c_i\approx 1$) when the prediction is accurate (i.e. $p_i\approx 1 \land y_i=1$ or $p_i\approx 0 \land y_i=0$) and the ensemble selector tends to select the predictive model rather than the ground-truth.
Otherwise, when the prediction error is large, the model should identify the prediction as uncertain (i.e. $c_i\approx 0$). 
In both cases, the combined prediction $p_i^\prime$ is always closer to the correct label.

Similar to Eq.\eqref{eq: quantile_prediction_loss}, the same substitution is adopted in the binary cross entropy learning objective and transforms it into:
\begin{equation}
    \mathcal{L}_\text{OR} = \sum_i \sum_t y_{i,t} \log p_{i,t}^\prime + (1-y_{i,t}) \log (1 - p_{i,t}^\prime)\label{eq: ordinal_regression_loss}
\end{equation}
By deriving the gradient of $p_i$ as:
\begin{equation}
\begin{split}
\frac{\partial \mathcal{L}_\text{OR}}{\partial p_{i,t}} &= y_{i,t}\frac{c_i}{p_{i,t}^\prime} + (y_{i,t}-1)\frac{c_i}{1-p_{i,t}^\prime} \\
&=\frac{y_{i,t}}{p_{i,t}+(\frac{1}{c_i} - 1)y_{i,t}}+\frac{y_{i,t} -1}{\frac{1-y_{i,t}}{c_i} - p_{i,t} + y_{i,t}}  
\end{split}
\end{equation}
we have the same conclusion as that in Eq.\eqref{eq: QR_p_gradient}: $p_i$ will learn faster when the model is confident (i.e. large $c_i$) and slower otherwise.

\subsection{Degradation of Uncontrolled Confidence}\label{sec: method_degradation}

Though the previous section shows how the watch-time prediction model optimizes under the control of the confidence model of $c$, the learning objectives of Eq.\eqref{eq: ordinal_regression_loss} and Eq.\eqref{eq: quantile_prediction_loss} do not restrict the optimization of $g_\phi$, which results in the degradation of model confidence.
Specifically, consider the gradient of $c_i$ for the two aforementioned objectives, we have:
\begin{equation}
\begin{aligned}
    \frac{\partial \mathcal{L}_\text{QP}}{\partial c_i} & = c_i (p_i-y_i)^2 \\
    \frac{\partial \mathcal{L}_\text{OR}}{\partial c_i} & = \frac{y_{i,t}}{\frac{1}{1-p_{i,t}} - c_i} + \frac{1-y_{i,t}}{\frac{1}{p_{i,t}} - c_i}\label{eq: gradient_c}
\end{aligned}
\end{equation}
We can directly see that $\partial \mathcal{L}_\text{QP}/\partial c_i \geq 0$ for all $c_i\in[0,1]$, and for both $y=0$ and $y=1$, $\partial \mathcal{L}_\text{OR}/\partial c_i \geq 0$ for $c_i\in[0,1],p_{i,t}\in(0,1)$.
This means that the confidence $c_i$ will continuously decrease as the training goes until it converges to zero, no matter how small the error is between $p$ and $y$.
Nevertheless, the gradient is larger when $c_i$ is larger for both gradients, which means that $c_i$ may quickly go to zero for all samples.
By the end of the story, the optimization reaches a shortcut $c_i\approx 0 \Rightarrow p^\prime\approx y$, which uses the ground-truth $y$ as the prediction, and the watch-time prediction model $p$ becomes reluctant to learn the error if no control is engaged upon the confidence model.

Unfortunately, the learning paradigm of Eq.\eqref{eq: ordinal_regression_loss} and Eq.\eqref{eq: quantile_prediction_loss} is inevitable since we need the information in the error between $p$ and $y$ to guide the learning of $c_i$.
We can observe this in the analysis of Eq.\eqref{eq: gradient_c}: both gradients are larger when $p$ has large errors (i.e., large $|p_i-y_i|$) and are smaller when the watch-time prediction is accurate.
This means that the confidence score will remain at a high level without significant change when the prediction is correct, and it will rapidly drop otherwise.

\subsection{Joint Optimization with Adversarial Confidence Maximization}\label{sec: method_joint_optimization}

\begin{figure}[t]
    \centering
    \includegraphics[width=0.9\linewidth]{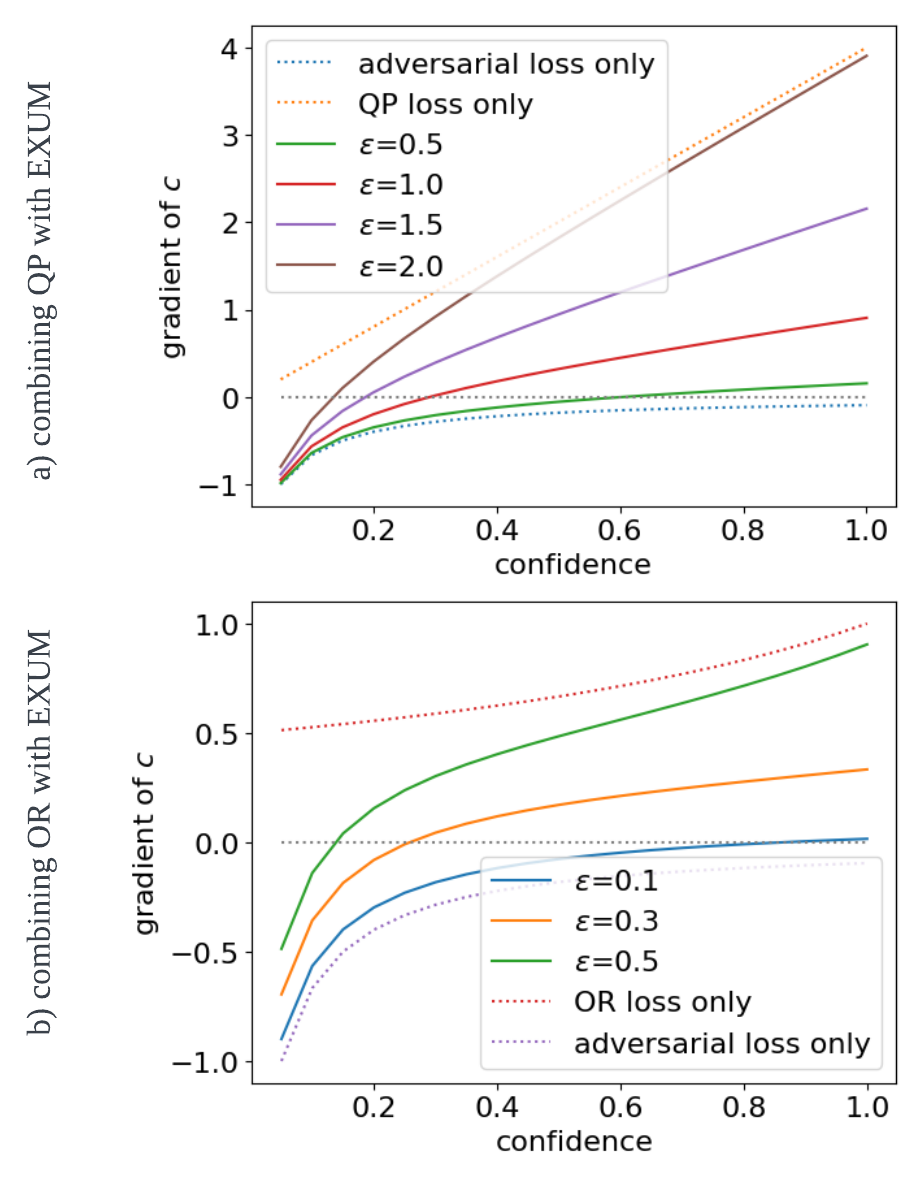}
    \caption{Gradient of $c$ and the theoretical converging point (with zero gradient) of EXUM.}
    \label{fig: combined_c_gradient}
\end{figure}

Inspired by the neural network confidence learning strategy in~\cite{devries2018learning} for the general out-of-distribution learning task, we propose to regulate $c_i$ through the following adversarial learning objective:
\begin{equation}
    \mathcal{L}_c = - \sum_i \log c_i\label{eq: c_adversarial}
\end{equation}
which aims to directly provide an opposite regulation against confidence degradation.
Semantically, this would actively ensure the increase of model confidence while learning accurate prediction.
Additionally, $\partial \mathcal{L}_c / \partial c_i$ is larger when $c_i$ is larger, which forms a reversed trend in the gradient magnitude compared to that in Eq.\eqref{eq: gradient_c}.
In other words, no matter how complicated the user's behavior is, the confidence model will always suppress the watch-time prediction model as much as possible to increase the model's overall confidence and reduce the error.

Finally, in order to accommodate different settings in practice, we formulate the joint learning objective of quantile prediction as:
\begin{equation}
    \mathcal{L}_\text{QP+EXUM} = \mathcal{L}_\text{QP} + \lambda \mathcal{L}_c\label{eq: combined_qp_loss}
\end{equation}
and that of the ordinal regression as:
\begin{equation}
    \mathcal{L}_\text{OR+EXUM} = \mathcal{L}_\text{OR} + \lambda \mathcal{L}_c\label{eq: combined_or_loss}
\end{equation}
where $\lambda$ controls the magnitude of adversarial confidence regulation.
We show the analysis of the gradient of $c$ conditioned on the same prediction error in Figure \ref{fig: combined_c_gradient}-a and Figure \ref{fig: combined_c_gradient}-b.
As we have discussed in section \ref{sec: method_degradation}, the original loss of Eq.\eqref{eq: ordinal_regression_loss} and Eq.\eqref{eq: quantile_prediction_loss} always generate a positive gradient causing the degradation of confidence, and the adversarial loss comes into rescue.
Specifically, represent the error term as $\epsilon=|p - y|$, then the combined gradient of Eq.\eqref{eq: combined_qp_loss} and that of Eq.\eqref{eq: combined_or_loss} are:
\begin{equation}
\begin{aligned}
    \frac{\partial \mathcal{L}_\text{QP+EXUM}}{\partial c_i} & = c_i \epsilon^2 - \frac{\lambda}{c_i}\\
    \frac{\partial \mathcal{L}_\text{OR+EXUM}}{\partial c_i} & = \frac{1}{\frac{1}{\epsilon} - c_i} - \frac{\lambda}{c_i}
    \label{eq: combined_gradient_qp_n_or}
\end{aligned}
\end{equation}

As shown in Figure \ref{fig: combined_c_gradient}, these gradients are monotonically increasing for both QP and OR modeling, so it intersect with the zero line in the middle of the confidence range rather than the two sides.
Note that the intersection points are $c_i^\ast=\sqrt{\lambda} / \epsilon$ for QP and $c_i^\ast=\lambda / (\epsilon (1+\lambda))$ for OR.
To ensure that the intersection occurs in the valid range $c_i^\ast\in[0,1]$, a careful design should be used for the choice of $\lambda$.
Specifically, for a small empirical error $\epsilon$, we should have $0<\lambda <\epsilon^2$ for QP, and $0<\lambda <\epsilon/(1-\epsilon)$ for OR.
In other words, we can choose a small positive $\lambda$ in order to make $c$ converge at some valid intersection point, so that it can represent the balance between the error-based confidence degradation and the intention of confidence promotion.

In practice, it is also possible to select a large $\lambda$ that is much greater than the error term, and the resulting optimization will follow the lower bound $-\lambda / c_i$ which is always negative.
This means that the confidence will monotonically increase until it converges to one.
In the view of the prediction model $p$, it will start from a relaxed confidence constraint which helps capture the uncertain user behavior, but it will gradually make more efforts to minimize the error to ensure the final prediction accuracy.
In general, we believe that the introduction of Eq.\eqref{eq: c_adversarial} should be effective as long as it provides sufficient adversarial forces against confidence degradation.
We will further illustrate this preferable feature in section \ref{sec: experiment_ablation}.



\section{Experiments}\label{sec: experiments}
As verification of our claims in this paper, we conduct experiments in both offline
and online environments.
As a guide for the remaining materials, we summarize the main research questions as follows:
\begin{itemize}
    \item \textbf{RQ1}(section \ref{sec: experiment_online} and section \ref{sec: experiment_offline_main}): Does EXUM improve the watch-time prediction accuracy while modeling the uncertainty/confidence of the prediction?
    \item \textbf{RQ2}(section \ref{sec: experiment_ablation}): Is EXUM sensitive to different backbones (i.e. quantile prediction vs. ordinal regression) or the hyperparameter $\lambda$?
    \item \textbf{RQ3}(section \ref{sec: experiment_confidence_analysis}): What is the behavior of the confidence model during training and what is its relationship to the prediction error during inference?
\end{itemize}

\subsection{Online A/B Test}\label{sec: experiment_online}

\begin{figure}[t]
\centering
\includegraphics[width=\linewidth]{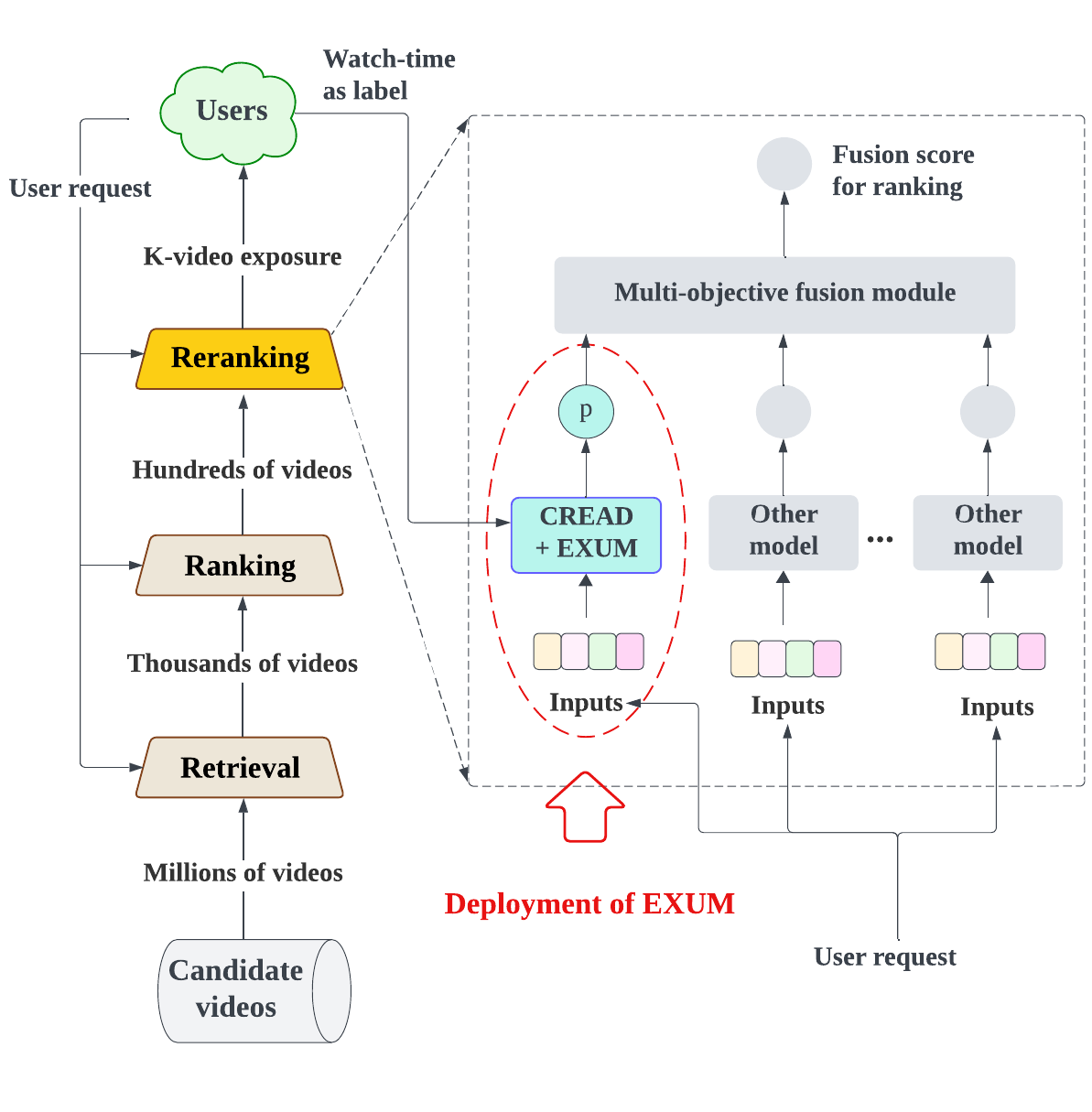}
\caption{Overview of online workflow.}
\label{fig: online_workflow}
\end{figure}

To validate the effectiveness of our approach in real-world scenarios, we conducted an online A/B test on an industrial video recommendation platform with more than 300 million active users and millions of video candidates every day.
Present the overview of the online recommender system as Figure \ref{fig: online_workflow}, where the system consists of a retrieval stage, a ranking stage, and a reranking stage.
The watch-time prediction task requires the exposure of videos so we choose to deploy EXUM in the final reranking stage, where it is used to improve an existing prediction model based on CREAD~\cite{sun2024cread}.
In addition to the watch-time prediction module, there also exist other scoring models (e.g. like-rate prediction) that collaboratively determine the final exposure of videos, forming an ensemble balanced by a multi-objective fusion module. Regarding the online deployment, we separate the embedding table maintainance (calling and update) into a separate service to improve the latency. Also note that the proposed EXUM framework only needs the watch-time prediction backbone without running the uncertainty prediction model during online inference. In terms of complexity, we need extra computation resources for the confidence prediction head (typically the same complexity as the watch-time prediction head) introduced during training, but no extra cost during inference. During the experiment, we randomly take 20\% traffic as the treatment group for 5 days to observe stable and statistically significant results, and compared it to the baseline CREAD already deployed in the rest of the traffic.

Notably, the predicted watch time is used as ranking scores, so the prediction error is no longer a direct evaluation.
Instead, we include the average \textbf{watch-time} (i.e. the amount of time a user stays on a video) as the main metric as the indicator of ranking performance, since more confident modeling of the watch-time should generate more accurate rankings and the top-K videos obtains more interactions.
In addition, modern video recommendation platforms also provide other functionalities such as ``like'' (i.e. clicking the like button), ``follow'' (i.e. subscribing to the video author), and ``comment'' (i.e. providing textual remarks)  for better social experiences. 
As a result, we also include the daily sum of \textbf{likes}, \textbf{follows}, and \textbf{comments} separately as additional metrics to observe the relations between watch-time and these interactions.
Intuitively, users have limited time so spending more time watching may indicate a corresponding drop in other interaction metrics.
In this sense, these interaction metrics serve as the constraint to indicate the stability of the model.

We report the daily performance comparison in Table \ref{tab: online_ab_results}. 
As we can see, the treatment group (CREAD+EXUM) achieved significant gains over $0.3\%$ in watch-time metric and the drop of constraint metrics is NOT statistically significant, i.e. like, follow, and comment drops within the range of variance (around 1\%). 
In contrast, the improvement in watch-time is statistically significant (i.e. $\geq 0.1\%$).
Additionally, the model quickly converges and the improvement in watch-time as well as the behavior of constraint metrics are consistent over the five-day observation period, indicating a stable online model learning and inference.
To further investigate how the EXUM model works and whether it can accommodate other watch-time prediction backbones, we conduct offline experiments as we will illustrate in the following sections.
\begin{table}[!htbp] 
    \centering
    \caption{Online A/B results.}
    \begin{tabular}{ccccccccccc} 
        \toprule 
        \multicolumn{3}{c}{\multirow{2}{*}{Days}}& \multicolumn{3}{c}{Main Metric}& &\multicolumn{3}{c}{Constraint Metrics}\\
        \multicolumn{3}{c}{}  & & Watch-Time & & &Like & Follow & Comment \\ 
        \hline
        \multicolumn{3}{c}{Day1}&   & 0.303\% &  & &-0.026\% & -0.247\% & 0.073\%\\ 
        \multicolumn{3}{c}{Day2}&  & 0.310\% &   & &-0.091\% & -0.363\% & -0.009\% \\
        \multicolumn{3}{c}{Day3}&  & 0.306\% & & &-0.073\%   & -0.343\% & -0.025\% \\
        \multicolumn{3}{c}{Day4}&  & 0.298\% & & & -0.105\%   & -0.209\% & -0.134\% \\
        \multicolumn{3}{c}{Day5}&  & 0.325\% & & &0.015\%   & -0.165\% & 0.001\% \\
        \bottomrule 
    \end{tabular}
    \label{tab: online_ab_results}
\end{table}

\subsection{Offline Experimental Setting}

\subsubsection{Datasets}

We conduct offline experiments on two public benchmark recommendation datasets including WeChat\footnote{https://algo.weixin.qq.com/} and KuaiRand\footnote{https://kuairand.com/} to provide a generalized analysis of the proposed method\footnote{Reproduction code can be found at https://anonymous.4open.science/r/EXUM/readme.md}.
Both datasets are collected from large micro-video recommendation platforms, i.e. WeChat Channels and Kuaishou:
\begin{enumerate}
    \item WeChat: This dataset was released by WeChat Big Data Challenge 2021, and it contains user interactions (including watch-time) on WeChat Channels which consists of 20000 users and 96418 items. 
    \item KuaiRand: This dataset is an unbiased sequential recommendation dataset collected from the recommendation logs of the video-sharing mobile app, Kuaishou. It consists of three datasets, KuaiRand-Pure, KuaiRand-1K, and KuaiRand-27K. We use the first two which contain 27077 users and 7551 items.
\end{enumerate}
For both datasets, we split the data into training (10 days in WeChat and 14 days in KuaiRand) and testing sets (2 days in WeChat and 10 days in KuaiRand) according to the dates of the samples, which mimic the chronological nature of the online environment.

\subsubsection{Methods and Model Specification} 
Here we list and specify the baselines and two applications of EXUM on distribution modeling methods:
\begin{itemize}
    \item \textbf{WLR\cite{covington2016deep}.} This method is firstly proposed to fit a weighted logistic regression model and use the watch time of the clicked samples as weight. 
    Then, use the learned odds as the predicted watch time. However, there are no clicked samples in our case, so we follow~\cite{zhan2022deconfounding} and adapt this method to our full-screen streaming scenario.
    \item \textbf{TPM\cite{lin2023tree}.} This method splits watch time into multiple ordinal intervals and iteratively merges the intervals into a balanced binary tree structure.
    The prediction is generated through a search process that is modeled as a sequence of binary decision-making problems.
    The training is based on a specialized progressive regression and allows adjustment through a confounding bias factor.
    \item \textbf{DML\cite{zhang2023leveraging}.} This method proposes a framework for relabeling and debiasing which transforms original user watch time into a range of training labels including Watch-time Percentile Rank (WPR), effective view (EV) and long view (LV). Specifically, WPR plays the greatest role and therefore is used for comparison in this paper.
    \item \textbf{D2Q\cite{zhan2022deconfounding}.} This approach splits data according to duration and fits a regression model to estimate watch time quantiles via mean squared error loss. 
    Then the predicted quantile is mapped to the watch-time value based on the empirical watch-time distribution. Details are presented in section \ref{sec: preliminary}.
    \item \textbf{CREAD\cite{sun2024cread}.} This approach is the ordinal regression approach mentioned in section \ref{sec: preliminary}.
    It discretizes the watch time into adaptive time segments, then uses a classification module to predict the watch time through multiple classification tasks, and finally uses a restoration module to output the watch time by multiplying the probability and the length of each time interval and summarizing.
    \item \textbf{D2Q-EXUM (Ours).} As described in Section \ref{sec: method}, this method is based on D2Q and adds the confidence prediction module with joint learning of Eq.\eqref{eq: combined_qp_loss}.
    The main task in this model is to estimate watch-time quantile via mean squared error loss, while the auxiliary task is to model the uncertainty via the adversarial loss of Eq.\eqref{eq: c_adversarial}.
    \item \textbf{CREAD-EXUM (Ours).} As described in Section \ref{sec: method}, this approach uses CREAD as the backbone and adds the confidence modeling module.
    The main watch-time prediction task aims to learn the probability of reaching each watch-time quantile and predict the final watch-time by the overall expectation.
    The auxiliary task aims to model the uncertainty with EXUM and the joint learning optimizes Eq.\eqref{eq: combined_or_loss}. 
\end{itemize}
For fair comparison, except for the output layers and loss functions, we adopt the same model structures as MLP with [128,64,32] as hidden dimensions, and both the watch-time prediction model and the confidence prediction model adopt MLP structure with [16] as hidden dimension. For all methods, we search the learning rate in the range [1e-5,1e-4,1e-3,1e-2] and select the settings with the best performance.

\subsubsection{Metrics and Evaluation Protocol}
For offline experiments, we include MAE the prediction error, and XAUC the ranking performance as indicators of the model's performance:
\begin{itemize}
    \item MAE (Mean Absolute Error): the conventional measurement for evaluating regression accuracy. 
    It measures the mean absolute error between the predicted and true values,
\[\text{MAE}=\sum_{i=1}^{N}\left | p_i - y_i  \right |  \]
where \(y_i\) is denoted as true watch-time and $p_i$ is denoted as the predicted watch-time, which is different from the definition in section \ref{sec: method}.
During comparison, a smaller value of MAE indicates a more accurate prediction.
    \item XAUC\cite{kallus2019fairness}: Note that an accurate regression performance does not necessarily indicate a better ranking performance, so we include this XAUC metric. 
    It first samples random instance pairs and checks whether their relative order is consistent with the ground truth. 
    Specifically, for a pair of samples, it will be scored 1 if the predicted watch-time values of the two videos are in the same order as the ground truth and scored 0 vice versa. 
    Finally, it takes the average scores as XAUC and larger values indicate a better ranking performance.
\end{itemize}

\begin{table}[h] 
    \centering
    \caption{Watch-time prediction results on KuaiRand and WeChat. Bold values denote the best performance, and underlined values denote the second best.}
    \begin{tabular}{ccccccccc} 
        \toprule 
        \multicolumn{3}{c}{\multirow{2}{*}{Methods}}& \multicolumn{2}{c}{KuaiRand}& &\multicolumn{2}{c}{WeChat}\\
        \multicolumn{3}{c}{}  & MAE$\downarrow$ & XAUC$\uparrow$  &  & MAE$\downarrow$ & XAUC$\uparrow$\\ 
        \midrule
        \multicolumn{3}{c}{WLR}& 25.1255 &0.6091 & & 12.5086 & 0.6324\\
        \multicolumn{3}{c}{TPM}& 24.2229 &0.6363 & & 12.2582  & 0.6376\\
        \multicolumn{3}{c}{DML}& \underline{21.1419} &\underline{0.6439} & & 11.9756  & 0.6607\\
        \midrule
        \multicolumn{3}{c}{D2Q}& 23.6171 &0.6268 & & \underline{11.3605} & \underline{0.6615}\\ 
        \multicolumn{3}{c}{D2Q-EXUM}& 23.3522 & 0.6349 & & \textbf{11.2351} & \textbf{0.6619} \\ 
        \multicolumn{3}{c}{Improvement}& 1.12\% & 1.29\% & & 1.10\%& 0.06\%\\ 
        \midrule
        \multicolumn{3}{c}{CREAD}& 22.2938 & 0.6375 & & 12.4797 &0.6341\\ 
        \multicolumn{3}{c}{CREAD-EXUM}& \textbf{18.5399} & \textbf{0.6526} & & 12.3723 & 0.6345\\ 
        \multicolumn{3}{c}{Improvement}& 16.84\% & 2.37\% & & 0.86\%& 0.06\%\\ 
        \bottomrule 
    \end{tabular}
    \label{tab: main_results}
\end{table}

\subsection{Main Results}\label{sec: experiment_offline_main}

We compare EXUM with current state-of-the-art methods on the two datasets.
We report the results in Table \ref{tab: main_results}. 
We can see that the distribution modeling baselines (i.e. D2Q and CREAD) are generally better than WLR and TPM, which verifies the correctness of stochastic behavior modeling.
When incorporating our proposed EXUM framework, D2Q-EXUM consistently outperforms its backbone D2Q over 1\% in MAE and XAUC on KuaiRand, and 1.1\% and 0.1\% in MAE and XAUC respectively on WeChat;
and CREAD-EXUM consistently outperforms its backbone CREAD by over 2\% in MAE and XAUC on KuaiRand, and 1\% in MAE on WeChat.
All these improvements are statistically significant (i.e. student t-test with $p<0.05$) except for a minor improvement of XAUC on WeChat when comparing CREAD-EXUM and CREAD.
In general, these results provide evidence that EXUM can effectively improve state-of-the-art distribution modeling methods in watch-time prediction tasks in both regression and ranking performances.

In addition, we notice that methods perform differently on the two datasets.
TPM can significantly improve the XAUC metric on KuaiRand dataset but this performance is not consistent on WeChat, indicating its unstable behavior.
In contrast, the consistent improvement of EXUM reveals its preferable generalization ability across different datasets and backbones.
Empirically, the MAE metrics of all methods are lower in WeChat while the ranking metrics are higher, compared to those in KuaiRand.
This indicates an easier watch-time prediction task in WeChat, which potentially explains why the improvement of EXUM is smaller in this dataset.
Besides, we also notice that the DML baseline (an enhanced version of D2Q) is slightly better than distribution modeling baselines in the KuaiRand dataset, but the performance is inconsistent in WeChat, indicating its data-dependent nature.
In any case, the best EXUM variants outperforms DML with significant improvements.



\subsection{Ablation}\label{sec: experiment_ablation}

To further investigate the behavior of the EXUM framework we conduct several ablation studies on the choices of backbones and the hyperparameters of the method.

\subsubsection{Distribution Confidence vs. Quantile Confidence:}
To integrate uncertainty modeling with OR for watch time prediction, the standard design in Eq.\eqref{eq: ordinal_regression_loss} applies the same confidence prediction $c_i$ across all watch-time prediction heads in the distribution.
In this design, the confidence model predicts the certainty of the entire predicted distribution.
Yet, we can extend the EXUM framework into each quantile prediction head and optimize the distribution at a more fine-grained level by predicting the confidence $c_{i,t}$ for each prediction head $p_{i,t}$.
The resulting ensemble becomes:
\begin{equation}
    p^\prime_{i,t} = c_{i,t} p_{i,t} + (1-c_{i,t}) y_{i,t} \label{eq: ordinal_prediction_per_head}
\end{equation}
We denote this alternative as CREAD-EXUM-MultiHead.
We conduct comparison experiments of the two alternatives (CREAD-EXUM and CREAD-EXUM-MultiHead) and summarize the results in Table \ref{tab: effect_of_confidence_head}.
For hyperparameters we select the best empirical settings with learning rate \(r=0.001\), epoch number \(E=20\), and adversarial loss weight \(\lambda=0.1\).
As we can see the original CREAD-EXUM method is slightly (i.e. not statistically significant) better than the multi-head alternative, indicating that the single-head confidence better expresses the certainty of the entire watch-time prediction model, and the separation of $c_i$ towards $c_{i,t}$ might be unnecessary.
Yet, both alternatives significantly improve the performance over the CREAD backbone, verifying the generalization ability of different model designs.

\begin{table}[!htbp] 
    \centering
    \caption{Comparison of the two alternatives of CREAD-EXUM on KuaiRand.}
    \begin{tabular}{ccccccccc} 
        \toprule 
        \multicolumn{3}{c}{\multirow{2}{*}{Design choice}}& \multicolumn{5}{c}{KuaiRand}& \\
        \multicolumn{3}{c}{}  & MAE$\downarrow$ & & & XAUC$\uparrow$  &  \\ 
        \hline  
        \multicolumn{3}{c}{CREAD}& 22.2938 & & & 0.6375  &  \\
        \multicolumn{3}{c}{CREAD-EXUM}& 18.5399 & & & 0.6526  &  \\  
        \multicolumn{3}{c}{CREAD-EXUM-MultiHead}& 19.0339 & & & 0.6507  &  \\
        \bottomrule 
    \end{tabular}
    \label{tab: effect_of_confidence_head}
\end{table}

\subsubsection{Effect of $\lambda$:}

\begin{figure}[h]
  \centering
  \includegraphics[scale=0.5]{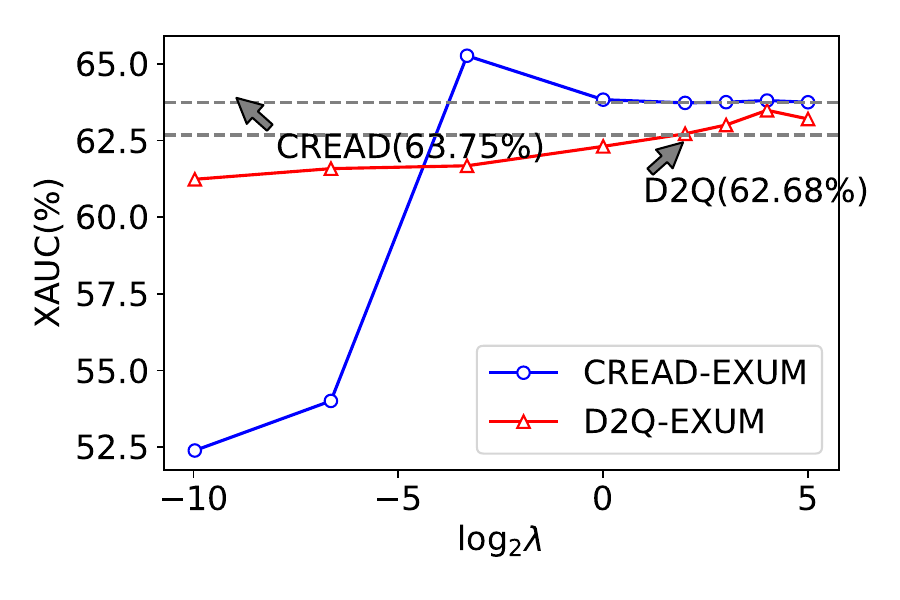}
  \caption{XAUC of watch-time prediction methods on KuaiRand with respect to confidence loss weight \(\lambda\). Best observed settings are $\lambda=0.1$ for CREAD and $\lambda=16$ for D2Q}
  \Description{Confidence loss weight.}
  \label{fig: confidence_loss_ablation}
\end{figure}

As we have discussed in section \ref{sec: method_joint_optimization}, key hyperparameter $\lambda$ controls the magnitude of the adversarial confidence learning, and consequently affects the final performance.
To analyze the model's sensitivity on $\lambda$, we conduct experiments on KuaiRand by altering this adversarial learning loss weight with \(\lambda \in \{0.001, 0.01, 0.1, 1.0, 4.0, 8.0, 16.0, 32.0\}\) and keeping other settings fixed (i.e. learning rate \(r=0.001\) and epoch number \(E=20\)).
To observe the effect on different backbones, we compare D2Q-EXUM, CREAD-EXUM, and their respective backbones.
We plot the XAUC results under different $\lambda$ as Figure \ref{fig: confidence_loss_ablation}, where the backbone results of D2Q and CREAD are presented as dotted horizontal lines.
As we can see, with an extremely small $\lambda$ (i.e. $\leq 0.01$ for CREAD and $\leq 1.0$ for D2Q), incorporating EXUM would reduce the model performance since the adversarial effect of Eq.\eqref{eq: c_adversarial} is not strong enough to avoid confidence degradation.
In these cases, the resulting prediction model becomes reluctant to learn the ground-truth label and the ensemble $p^\prime$ tends to find the shortcut that always predicts $y$ with itself.
In comparison, when increasing $\lambda$ to a sufficiently large value, EXUM generates superior performance over the backbones for both D2Q and CREAD.
However, the performance may also gradually decrease when further increasing $\lambda$ to larger values.
In these cases, the adversarial learning dominates the updates of $c$ and the confidence quickly converges to one, and the ensemble model degrades into $p^\prime=p$.
In summary, there exists an optimal choice of $\lambda$ in the range $[0,\infty)$, and its value depends on the distribution modeling backbone.





\textbf{Quantile Prediction vs. Ordinal Regression}

As presented by Table \ref{tab: main_results}, the QP methods (i.e. D2Q and D2Q-EXUM) outperform the OR methods (i.e. CREAD and CREAD-EXUM) in WeChat but they perform worse than OR methods in KuaiRand.
In other words, neither QP nor OR is currently a universally superior solution in the field and the performance depends on the data characteristics.
Besides, the improvement of EXUM is not large enough to surpass the gap between the two backbones, indicating the importance of backbone selection in different environments.
In our industrial experiments mentioned in section \ref{sec: experiment_online}, we found the CREAD backbone superior to D2Q.
In addition, as we can see in Figure \ref{fig: confidence_loss_ablation}, CREAD is more sensitive to $\lambda$ compared to D2Q when $\lambda<1$, but can achieve better results as long as $\lambda$ is sufficiently large.
This may indicate that QP methods are more stable in learning but also may not always explore the optimal performance.

\begin{figure}[ht]
    \centering
    \includegraphics[scale=0.5]{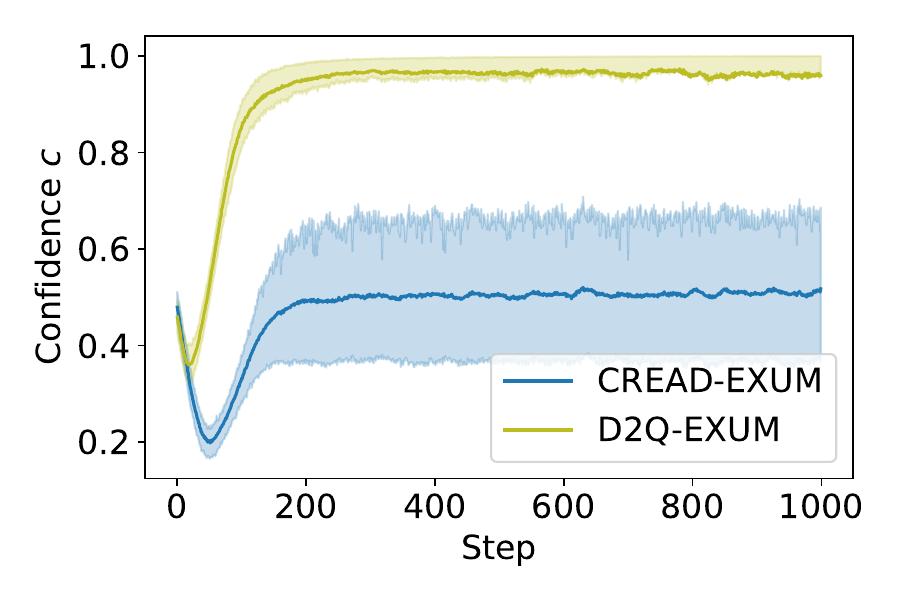}
    \caption{Confidence $c$ in the training process on KuaiRand where \(\lambda_{D2Q-EXUM} = 16.0\) and \(\lambda_{CREAD-EXUM} = 0.1\).}
    \Description{Confidence training.}
    \label{fig: c_curve}
\end{figure}

\subsection{Analysis on Confidence Module}\label{sec: experiment_confidence_analysis}

To better understand the functionalities of the confidence module, we select the best setting for D2Q-EXUM (with $\lambda=16.0$) and CREAD-EXUM (with $\lambda=0.1$) and observe the training curves of $c$.
As shown in Figure \ref{fig: c_curve}, the solid lines represent the mean value of $c$ of all samples, while the shaded areas represent the variance. For both methods, the confidence first drops as Eq.\eqref{eq: combined_gradient_qp_n_or} is dominated by the large error terms and relatively small $c$, indicating an initial tendency of confidence degradation.
Then, as the learning continues, the gradient information from the adversarial learning gradually surpasses that from the QP loss (or OR loss) since the error $\epsilon$ drops.
This results in a re-bounce of $c$ into higher confidence values, indicating the effectiveness of the adversarial regularization.
Finally, the value of $c$ converges in two circumstances:
1) As in D2Q-EXUM, when the optimal $\lambda=16.0$ is much larger than $c$, its optimization of the adversarial loss in Eq.\eqref{eq: c_adversarial} will dominate the updates of $c$ and results in a convergences towards $c=1$, which verifies our analysis in section \ref{sec: method_joint_optimization}.
2) As in CREAD-EXUM, when the optimal $\lambda=0.1$ is relatively small and possibly with the range $[0,\epsilon^2]$, the converging point of $c\approx 0.5$ occurs in the middle of the range $[0,1]$, which verifies the existence of the balance between the error-based confidence degradation and the intention of confidence promotion, as we have mentioned in section \ref{sec: method_joint_optimization}. Furthermore, it can be seen that the variance remains stable after convergence and method with higher mean confidence (i.e. D2Q) has a reduced confidence variance about its prediction while sample differences are higher for the model with lower mean confidence (i.e. CREAD).
In general, we believe that the confidence model is related to the prediction error but not necessarily in a linear relation, and it is still an open question whether there exists an optimal way to express the uncertainty of the watch-time prediction model under arbitrary distribution.

\section{Conclusion}\label{sec: conclusion}

To accurately estimate the watch time in video recommender systems, it is critical to capture the stochastic user behavior with distribution modeling techniques.
In this paper, we propose the EXUM framework that can explicitly model the uncertainty of the predicted watch-time distribution on both quantile prediction and ordinal regression backbones.
We show that the resulting framework needs a joint optimization of error minimization and confidence promotion.
The empirical study in online A/B testing and offline evaluation verifies the superiority of EXUM.
We also show that the introduced confidence model can reach an adequate converging point with carefully selected loss balancing factor $\lambda$.
In practice, we found CREAD-EXUM performs the best in more sophisticated environments with complicated user behaviors, but we have provided empirical evidence that the EXUM framework is likely to generalize to different backbone models across various datasets.

\newpage
\balance
\bibliographystyle{ACM-Reference-Format}
\bibliography{main}

\end{document}